# Enhancing super-resolution ultrasound localisation through multi-frame deconvolution exploiting spatiotemporal coherence

Su Yan[a], Clotilde Vié[a], Marcelo Lerendegui[a], Herman Verinaz-Jadan[b], Jipeng Yan[a], Martina Tashkova[c], James Burn[d], Bingxue Wang[a], Gary Frost[c], Kevin G. Murphy[c], Meng-Xing Tang[a,*]

[a] *Ultrasound Lab for Imaging and Sensing, the Department of Bioengineering, Faculty of Engineering, Imperial College London, London, UK.*
[b] *Faculty of Electrical and Computer Engineering, Escuela Superior Politécnica del Litoral (ESPOL), Guayaquil, Ecuador.*
[c] *The Department of Metabolism, Digestion and Reproduction, Faculty of Medicine, Imperial College London, London, UK.*
[d] *Imperial College Healthcare NHS Trust, London, UK*

## Abstract

Super-resolution ultrasound imaging through microbubble (MB) localisation and tracking, also known as ultrasound localisation microscopy, allows non-invasive sub-diffraction resolution imaging of microvasculature in animals and humans. The number of MBs localised from the acquired contrast- enhanced ultrasound (CEUS) images and the localisation precision directly influence the quality of the resulting super-resolution microvasculature images. However, non-negligible noise present in the CEUS images can make localising MBs challenging. To enhance the MB localisation performance, we propose a Multi-Frame Deconvolution (MF-Decon) framework that can exploit the spatiotemporal coherence inherent in the CEUS data, with new spatial and temporal regularisers designed based on total variation (TV) and regularisation by denoising (RED). Based on the MF-Decon framework, we introduce two novel methods: MF-Decon with spatial and temporal TVs (MF-Decon+3DTV) and MF-Decon with spatial RED and temporal TV (MF-Decon+RED+TV). Results from in silico simulations indicate that our methods outperform two widely used methods using deconvolution or normalised cross-correlation across all evaluation metrics, including precision, recall, $F_1$ score, mean and standard localisation errors. In particular, our methods improve MB localisation precision by up to 39% and recall by up to 12%. Super-resolution microvasculature maps generated with our methods on a publicly available in vivo rat brain dataset show less noise, better contrast, higher resolution and more vessel structures.

*Corresponding author. Tel.: +44-2075943664
*Email address:* mengxing.tang@imperial.ac.uk (Meng-Xing Tang)

# 1. Introduction

Super-resolution ultrasound (SRUS) imaging through localisation and tracking of contrast agents such as microbubbles (MBs), also known as ultrasound localization microscopy (ULM), is capable of imaging the microvasculature in animals and humans beyond the wave diffraction limit (Christensen-Jeffries et al., 2020). Successful applications of SRUS/ULM have been demonstrated in various contexts, including in vitro (Desailly et al., 2013; Viessmann et al., 2013), in vivo (Christensen-Jeffries et al., 2015; Errico et al., 2015; Ackermann and Schmitz, 2016; Song et al., 2018; Bar-Zion et al., 2018; Couture et al., 2018; Zhu et al., 2019; Van Sloun et al., 2020; Demeulenaere et al., 2022; Renaudin et al., 2022; Taghavi et al., 2022; Wu et al., 2024), and in clinical settings (Opacic et al., 2018; Huang et al., 2021; Demené et al., 2021; Zhu et al., 2022; Yan et al., 2024).

Accurate localisation of MBs in contrast-enhanced ultrasound (CEUS) images is crucial for generating high-quality super-resolution (SR) microvasculature maps. The number of MBs localised and the precision of their localisation directly influence the image quality. Typically, a long acquisition time is required to ensure sufficient MBs have passed through the vessels of the target organ, making them visible in the final SR maps. However, long acquisition times are often associated with increased organ motion. To mitigate this, MB concentration is usually increased, allowing more MBs to flow into the target organ within a shorter acquisition time. This approach, however, can cause overlapping MB signals that are difficult to separate. More importantly, non-negligible noise and artefacts in the acquired CEUS images present a major challenge, especially in vivo. This can cause errors in MB localisation, resulting in noisy SR microvasculature maps with lost or incorrect vessel structures. Therefore, there is a need for a robust method capable of effectively and accurately localising MBs from noisy CEUS images for SRUS imaging.

Several methods have been proposed to address the problem of MB localisation in SRUS imaging. Techniques such as peak detection and centroid detection methods are employed to localise the MBs from the CEUS images after background noise removal (Christensen-Jeffries et al., 2015; Heiles et al., 2022). However, they are also prone to mislabelling high-intensity noise as MBs. Normalised cross-correlation (NCC) with a point spread function (PSF) of the MB is applied to counter noise (Song et al., 2018; J. Yan et al., 2023), but its ability to separate nearby MBs is limited. To better isolate overlapping MBs, researchers have modelled MB localisation as a sparsity-based image deconvolution problem. This problem is solved with an approximated PSF with methods like the Richardson-Lucy algorithm (Chen et al., 2020; Qian et al., 2020), or the fast iterative shrinkage-thresholding algorithm (FISTA) (Bar-Zion et al., 2018; Solomon et al., 2019; Yan et al., 2022). These deconvolution methods can separate overlapping MBs better, but noise can still greatly reduce their performance. The development of deep learning and neural networks has supported the use of network-based approaches for SRUS (Liu et al., 2020; Van Sloun et al., 2020; Milecki et al., 2021; Van Sloun et al., 2021; Chen et al., 2022; You et al., 2023; Shin et al., 2024; Rauby et al., 2024). Although neural networks can localise MBs at high concentrations, their performance varies greatly with training datasets. This variation causes problems in adapting the networks to in vivo and clinical cases, where the acquired data are limited and no ground truth is available. Moreover, the limited generalisation ability of the networks makes them less reliable, particularly in clinical applications.

Since the CEUS data from a single acquisition forms a three-dimensional (3D) spatiotemporal tensor, it is possible to leverage both spatial and temporal coherence in processing the data. Although only a few papers have considered this coherence in MB localisation, especially the coherence in the temporal dimension, the results suggest that it is a promising direction worth exploring. For example, flow kinematics of individual MBs have been used as additional sparsity weights in deconvolution (Solomon et al., 2019). More recently, a Track-and-Localise workflow in (Leconte et al., 2023) has been proposed to enhance results by first tracking the trajectories of MBs and then localising them. Similarly, the LOCA-ULM network (Shin et al., 2024) localises MBs based on three adjacent frames. Building on existing research, we introduce a novel Multi-Frame sparsity-based Deconvolution (MF-Decon) framework designed to exploit spatiotemporal coherence for localising MBs in noisy CEUS images. In contrast to conventional methods that process individual frames independently, our proposed MF-Decon framework deconvolves the multi-frame CEUS data from a single acquisition as a whole. We address the problem in three dimensions by incorporating regularisation techniques across the two spatial dimensions and one temporal dimension. These regularisers, based on total variation (TV) (Rudin et al., 1992) and regularisation by denoising (RED) (Romano et al., 2017), are specifically designed to enhance spatiotemporal coherence during deconvolution. Based on these developments, this paper proposes two novel methods for MB localisation in SRUS imaging: MF-Decon with spatial and temporal TVs (MF-Decon+3DTV) and MF-Decon with spatial RED and temporal TV (MF-Decon+RED+TV). Finally, the performance of the MF-Decon+3DTV and MF-Decon+RED+TV methods is evaluated using both in silico simulations and an in vivo rat brain dataset, with comparisons to the existing widely used MB localisation methods.

# 2. Methods

In this section, we begin by reviewing the traditional deconvolution (Decon) method. Based on this, we then introduce our proposed Multi-Frame Deconvolution (MF-Decon) framework. Subsequently, we dive into the details of the proposed MF-Decon+3DTV and MF-Decon+RED+TV methods, detailing the associated optimisation schemes for the constrained multi-frame deconvolution problems. Finally, we explain how the performances of the proposed methods are evaluated.

## 2.1. Traditional deconvolution

Image deconvolution can help localise MBs from the ultrasound images, by shrinking individual MBs, and reduce background noises. Let $Y \in \mathbb{R}^{W \times H}$ denote a two-dimensional (2D) CEUS image that has been motion-corrected and tissue-removed, where $W$ and $H$ denote the width and height in pixels, respectively. Given the PSF of the imaging system, represented by $A \in \mathbb{R}^{w \times h}$, the acquired CEUS image can be modelled as a convolution of the PSF and a high-resolution version of the bubble image $X \in \mathbb{R}^{W \times H}$ plus noise $X \in \mathbb{R}^{W \times H}$ (Bar-Zion et al., 2018; Demene et al., 2015; Yan et al., 2022):

$$Y = A * X + E, \tag{1}$$

where $*$ represents the multi-dimensional convolution operation. To estimate the PSF $A$, several separated MB signals are manually selected from the CEUS images, and their average shape is used. Specifically, we estimate the PSF by fitting a 2D Gaussian to the average of 10 isolated MBs manually selected for each acquisition. The noise in beamformed ultrasound images is known to follow a Rician distribution (Shin et al., 2024), which can be well approximated by a Gaussian distribution even when the signal-to-noise-ratio (SNR) is relatively low (Gudbjartsson and Patz, 1995). Given that the noise can be assumed to be Gaussian, it is appropriate to use mean squared error minimisation in the deconvolution approach. Therefore, image deconvolution aims to retrieve the high-resolution deconvolved image, enabling more precise localisation of overlapped MBs by solving the inverse problem (Yan et al., 2022):

$$\hat{X} = \underset{X}{\operatorname{argmin}} \frac{1}{2} \|Y - A * X\|_F^2. \tag{2}$$

Given the typically low concentration and sparse distribution of MBs, sparsity is imposed on problem (2) by incorporating an $\ell_1$ norm regularisation term (Bar-Zion et al., 2018). Additionally, a non-negativity constraint ensures that all MB intensity values remain positive. Consequently, problem (2) is reformulated as:

$$\hat{X} = \underset{X}{\operatorname{argmin}} \frac{1}{2} \|Y - A * X\|_F^2 + \mathcal{I}_+(X) + \lambda_1 \|X\|_1, \tag{3}$$

where $\mathcal{I}_+(\cdot)$ denotes the non-negativity constraint and $\|\cdot\|_1$ denotes the entrywise $\ell_1$ norm. In this paper, unless stated otherwise, we are specifically using entrywise norms. This is a classic deconvolution problem and can be solved using gradient-descent-based optimisation methods, such as the fast iterative shrinkage-thresholding algorithm (FISTA) (Beck and Teboulle, 2009) and the Alternating Direction Method of Multipliers (ADMM) (Boyd et al., 2011). After deconvolution, a noise threshold is applied to binarise the deconvolved frame into a binary location mask. Finally, the MATLAB built-in weighted centroid detection function 'regionprops' is used to determine the centre of each MB based on the binary mask and the intensities in the deconvolved image. The framework of the traditional deconvolution (Decon) method is demonstrated in Fig. 1(a).

## 2.2. Multi-frame deconvolution framework

To generate SR microvasculature images, a series of consecutive CEUS images are acquired, totalling $K$ frames. Assuming that the PSF remains constant across the frames, the model in (1) still holds for multiple frames:

$$\mathbf{Y} = A * \mathbf{X} + \mathbf{E}, \tag{4}$$

where $\mathbf{Y} \in \mathbb{R}^{W \times H \times K}$ represents the multi-frame CEUS images, $\mathbf{X} \in \mathbb{R}^{W \times H \times K}$ represents the multi-frame high-resolution deconvolved images and $\mathbf{E} \in \mathbb{R}^{W \times H \times K}$ is the noise. Importantly, the convolution operation $*$ is applied slice-wise, meaning that each 2D slice of the 3D matrix ($\mathbf{X}$) is independently convolved with ($A$).

Instead of processing the multi-frame CEUS images frame by frame traditionally, we can deconvolve the entire data across all frames at the same time using the same non-negativity and sparsity:

$$\widehat{\mathbf{X}} = \underset{\mathbf{X}}{\operatorname{argmin}} \frac{1}{2} \|\mathbf{Y} - A * \mathbf{X}\|_F^2 + \mathcal{I}_+(\mathbf{X}) + \lambda_1 \|\mathbf{X}\|_1, \tag{5}$$

The multi-frame deconvolution problem in (5) can also be solved with FISTA or ADMM.

To further reduce noise, we propose a multi-frame strategy for setting the noise threshold after deconvolution. An averaged noise image is created by averaging all the noise images estimated using an adaptive image threshold technique (Bradley and Roth, 2007) across all frames. After that, the deconvolved frames are binarised by thresholding based on this averaged noise image, and the centres of MBs are also retrieved using Matlab centroid detection function 'regionprops'. Finally, the proposed Multi-Frame Deconvolution (MF-Decon) framework is summarised in Fig. 1(b).

## 2.3. MF-Decon with spatiotemporal coherence

In this section, we introduce the regularisation techniques incorporated into the MF-Decon problem in (5), designed to exploit the spatiotemporal coherence of multi-frame CEUS images and thereby enhance the localisation performance.

Two-dimensional total variation (TV) is a spatial regularisation technique that can effectively remove noise while simultaneously preserving the genuine features of images (Rudin et al., 1992). We apply the 2D TV regulariser on the reconstructed MB images ($A * \mathbf{X}$) to reduce noise. As an alternative, Regularisation by denoising (RED), introduced more recently in (Romano et al., 2017), is another powerful framework that has also been applied to various problems in ultrasound imaging (Goudarzi et al., 2022). RED allows the use of an arbitrary denoiser as a regulariser for inverse problems in image processing, such as image denoising, image deblurring and image super-resolution. Our framework supports either regularisation technique, allowing for the substitution of 2D TV with RED based on specific requirements. More importantly, as the multi-frame CEUS images are acquired sequentially, they contain temporal information often overlooked in traditional localisation and deconvolution methods. In the time dimension, the behaviour of MBs is distinctly captured. A fast-moving MB causes a signal at a single pixel to quickly rise from baseline to a peak and then rapidly return, forming a sharp, intense peak. Conversely, a slow-moving MB produces a more gradual increase and decrease in signal intensity, resulting in a wider peak. Noise, which typically lacks temporal coherence and fluctuates more rapidly than MB movements, results in frequent, erratic changes in the signal. Removing the noise while keeping the MB signals is effectively equivalent to reducing the rapid temporal fluctuations while preserving temporal coherent features, aligning with the principles of TV regularisation (Rudin et al., 1992). Therefore, we incorporate the spatiotemporal coherence of the CEUS images into the MF-Decon framework with the following formulation:

$$\begin{aligned}\hat{\mathbf{X}} = \underset{\mathbf{X}}{\operatorname{argmin}} \frac{1}{2} \|\mathbf{Y} - A * \mathbf{X}\|_F^2 + \mathcal{I}_+(\mathbf{X}) + \lambda_1 \|\mathbf{X}\|_1 \\ + \lambda_2 \mathcal{C}_s(A * \mathbf{X}) + \lambda_3 \mathcal{C}_t(A * \mathbf{X}),\end{aligned} \tag{6}$$

The spatial regulariser $\mathcal{C}_s(A * \mathbf{X})$ is proposed based on the 2D TV or RED:

$$\mathcal{C}_s(A * \mathbf{X}) = \begin{cases} TV_{2D}(A * \mathbf{X}), & \text{in case of 2D TV,} \\ RED(A * \mathbf{X}), & \text{in case of RED,} \end{cases} \tag{7}$$

and the temporal regulariser $\mathcal{C}_t(A * \mathbf{X})$ is based on the 1D TV in the time dimension:

$$\mathcal{C}_s(A * \mathbf{X}) = TV_{1D}(A * \mathbf{X}). \tag{8}$$

In the next two sub-sections, we introduce the details of our proposed MF-Decon methods: MF-Decon with Spatial and Temporal TV (MF-Decon+3DTV) and MF-Decon with Spatial RED and Temporal TV (MF-Decon+RED+TV).

#### 2.3.1. MF-Decon with spatial and temporal TV

We first address the MF-Decon problem in (6) with spatial and temporal TV (MF-Decon+3DTV), which will serve as the basis for the solution with spatial RED and temporal TV. Problem (6) can now be expressed as:

$$\begin{aligned}\hat{\mathbf{X}} = \underset{\mathbf{X}}{\operatorname{argmin}} \frac{1}{2} \|\mathbf{Y} - A * \mathbf{X}\|_F^2 + \mathcal{I}_+(\mathbf{X}) + \lambda_1 \|\mathbf{X}\|_1 \\ + \lambda_2 \|\mathbf{d}_1 * (A * \mathbf{X})\|_1 + \lambda_2 \|\mathbf{d}_2 * (A * \mathbf{X})\|_1 + \lambda_3 \|\mathbf{d}_3 * (A * \mathbf{X})\|_1,\end{aligned} \tag{9}$$

where $\mathbf{d}_1 \in \mathbb{R}^{3\times1\times1}$, $\mathbf{d}_2 \in \mathbb{R}^{1\times3\times1}$, and $\mathbf{d}_3 \in \mathbb{R}^{1\times1\times3}$ are the convolutional kernels that compute the derivatives along the 1st, 2nd, and 3rd dimensions of a tensor, given by:

$$\mathbf{d}_{1_{:,1,1}} = \mathbf{d}_{2_{1,:,1}} = \mathbf{d}_{3_{1,1,:}} = [0,1,-1]. \tag{10}$$

In this formulation, $\mathbf{d}_1$ and $\mathbf{d}_2$ are utilised to enforce 2D TV regularisation on the spatial dimensions, while $\mathbf{d}_3$ is used to impose 1D TV regularisation on the temporal dimension. We can reformulate this optimisation problem to make it suitable for ADMM (Boyd et al., 2011) by expressing it in an equivalent form with four auxiliary variables $\mathbf{Z}_1$, $\mathbf{Z}_2$, $\mathbf{Z}_3$, and $\mathbf{Z}_4$:

$$\begin{aligned}\hat{\mathbf{X}} = \underset{\mathbf{X}}{\operatorname{argmin}} \frac{1}{2} \|\mathbf{Y} - A * \mathbf{X}\|_F^2 + \mathcal{I}_+(\mathbf{Z}_1) + \lambda_1 \|\mathbf{Z}_1\|_1 \\ + \lambda_2 \|\mathbf{Z}_2\|_1 + \lambda_2 \|\mathbf{Z}_3\|_1 + \lambda_3 \|\mathbf{Z}_4\|_1 \\ \text{s.t.} \quad \mathbf{Z}_1 = \mathbf{X}, \mathbf{Z}_2 = \mathbf{d}_1 * (A * \mathbf{X}), \\ \mathbf{Z}_3 = \mathbf{d}_2 * (A * \mathbf{X}), \mathbf{Z}_4 = \mathbf{d}_3 * (A * \mathbf{X}).\end{aligned} \tag{11}$$

**Algorithm 1:** MF-Decon with spatial and temporal TV (MF-Decon+3DTV) solved with ILF-ADMM

**Input:** Multi-frame CEUS images $\mathbf{Y}$ and estimated PSF $A$

**Output:** High-resolution deconvolved images $\widehat{\mathbf{X}}$

Initialise matrices $\mathbf{X}^{(0)} = \mathbf{0}$, ${\mathbf{Z}_1}^{(0)} = \mathbf{0}$, ${\mathbf{Z}_2}^{(0)} = \mathbf{0}$, ${\mathbf{Z}_3}^{(0)} = \mathbf{0}$, ${\mathbf{Z}_4}^{(0)} = \mathbf{0}$, ${\tilde{\mathbf{Z}}_1}^{(0)} = \mathbf{0}$, ${\tilde{\mathbf{Z}}_2}^{(0)} = \mathbf{0}$, ${\tilde{\mathbf{Z}}_3}^{(0)} = \mathbf{0}$, ${\tilde{\mathbf{Z}}_4}^{(0)} = \mathbf{0}$

$m = 1$

**while** *not converged* **do**

$\quad {\hat{\mathbf{Z}}_1}^{(m)} = \mathrm{prox}_{\mathcal{J}_+}\left(\mathbf{X}^{(m-1)} + {\tilde{\mathbf{Z}}_1}^{(m-1)}\right)$

$\quad {\mathbf{Z}_1}^{(m)} = \mathrm{prox}_{\|\cdot\|_{1,\lambda_1}}\left({\hat{\mathbf{Z}}_1}^{(m)}\right)$

$\quad {\mathbf{Z}_2}^{(m)} = \mathrm{prox}_{\|\cdot\|_{1,\lambda_2}}\left(\mathbf{d}_1 * \left(A * \mathbf{X}^{(m-1)}\right) + {\tilde{\mathbf{Z}}_2}^{(m-1)}\right)$

$\quad {\mathbf{Z}_3}^{(m)} = \mathrm{prox}_{\|\cdot\|_{1,\lambda_2}}\left(\mathbf{d}_2 * \left(A * \mathbf{X}^{(m-1)}\right) + {\tilde{\mathbf{Z}}_3}^{(m-1)}\right)$

$\quad {\mathbf{Z}_4}^{(m)} = \mathrm{prox}_{\|\cdot\|_{1,\lambda_3}}\left(\mathbf{d}_3 * \left(A * \mathbf{X}^{(m-1)}\right) + {\tilde{\mathbf{Z}}_4}^{(m-1)}\right)$

$\quad$ Update $\mathbf{X}^{(m)}$ according to (14)

$\quad {\tilde{\mathbf{Z}}_1}^{(m)} = {\tilde{\mathbf{Z}}_1}^{(m-1)} + \mathbf{X}^{(m)} - {\mathbf{Z}_1}^{(m)}$

$\quad {\tilde{\mathbf{Z}}_2}^{(m)} = {\tilde{\mathbf{Z}}_2}^{(m-1)} + \mathbf{d}_1 * \left(A * \mathbf{X}^{(m)}\right) - {\mathbf{Z}_2}^{(m)}$

$\quad {\tilde{\mathbf{Z}}_3}^{(m)} = {\tilde{\mathbf{Z}}_3}^{(m-1)} + \mathbf{d}_2 * \left(A * \mathbf{X}^{(m)}\right) - {\mathbf{Z}_3}^{(m)}$

$\quad {\tilde{\mathbf{Z}}_4}^{(m)} = {\tilde{\mathbf{Z}}_4}^{(m-1)} + \mathbf{d}_3 * \left(A * \mathbf{X}^{(m)}\right) - {\mathbf{Z}_4}^{(m)}$

$\quad m = m + 1$

$\widehat{\mathbf{X}} = \mathbf{X}^{(m)}$

In this paper, we apply a variation of ADMM, an inner-loop free ADMM (ILF-ADMM) (Donati et al., 2019), which has been demonstrated effective for similar regularised optimisation problems (Verinaz-Jadan et al., 2022; S. Yan et al., 2023). ILF-ADMM was proposed to avoid the inner gradient descent loops in the optimisation and thus have a faster convergence. The full details of the optimisation scheme for (11) are demonstrated in Algorithm 1, where $\tilde{\mathbf{Z}}_1$, $\tilde{\mathbf{Z}}_2$, $\tilde{\mathbf{Z}}_3$ and $\tilde{\mathbf{Z}}_4$ are dual variables added for ADMM. Moreover, $(\cdot)'$ is the operation that flips a tensor along all dimensions, $\mathrm{prox}_{\mathcal{J}_+}$ is the proximal operator that enforces the non-negativity constraint:

$$\mathrm{prox}_{\mathcal{J}_+}(x) = \begin{cases} x, & \text{if } x \geq 0, \\ 0, & \text{if } x < 0, \end{cases} \tag{12}$$

and $\mathrm{prox}_{\|\cdot\|_{1,\lambda_1}}$ is the proximal operator that promotes sparsity in the estimated $\mathbf{X}$ through regularisation, which is equivalent to an element-wise soft-thresholding operation with a threshold value $\lambda_1$:

$$\mathrm{prox}_{\|\cdot\|_{1,\lambda_1}}(x) = \begin{cases} x - \lambda_1, & \text{if } x > \lambda_1, \\ 0, & \text{if } |x| \leq \lambda_1, \\ x + \lambda_1, & \text{if } x < -\lambda_1. \end{cases} \tag{13}$$

The term $\mathbf{h}$ is added to avoid the gradient decent loop inside the traditional ADMM and achieve faster convergence, given by $\mathbf{h} = \alpha\|A\|^2\boldsymbol{\delta}$, where $\boldsymbol{\delta}$ is a 3D Kronecker delta and $\alpha$ is the learning rate. According to (Donati et al., 2019), $\mathbf{X}^{(m)}$ can be updated following:

$$\begin{aligned}(\rho_1\boldsymbol{\delta} + \rho_2\mathbf{d}_1' * A' * A * \mathbf{d}_1 + \rho_2\mathbf{d}_2' * A' * A * \mathbf{d}_2 \\ +\rho_3\mathbf{d}_3' * A' * A * \mathbf{d}_3 + \mathbf{h}) * \mathbf{X}^{(m)} = \mathbf{W}^{(m)},\end{aligned} \tag{14}$$

with

$$\begin{aligned}\mathbf{W}^{(m)} = A' * \mathbf{Y} + \rho_1\left(\mathbf{Z}_1{}^{(m)} - \tilde{\mathbf{Z}}_1{}^{(m-1)}\right) + \rho_2\mathbf{d}_1' * \left(A' * \left(\mathbf{Z}_2{}^{(m)} - \tilde{\mathbf{Z}}_2{}^{(m-1)}\right)\right) \\ +\rho_2\mathbf{d}_2' * \left(A' * \left(\mathbf{Z}_3{}^{(m)} - \tilde{\mathbf{Z}}_3{}^{(m-1)}\right)\right) + \rho_3\mathbf{d}_3' * \left(A' * \left(\mathbf{Z}_4{}^{(m)} - \tilde{\mathbf{Z}}_4{}^{(m-1)}\right)\right) \\ +(\mathbf{h} - A' * A) * \mathbf{X}^{(m-1)},\end{aligned} \tag{15}$$

where $\rho_1$, $\rho_2$ and $\rho_3$ are coefficients to control the importance weights of different constraints in ADMM. Since convolution corresponds to element-wise multiplication in the Fourier domain, a more straightforward method to solve (14) is as follows:

$$\begin{aligned}\mathbf{X}^{(m)} = \mathcal{F}^{-1}\{\mathcal{F}(\mathbf{W}^{(m)})/[\,\rho_1 + \rho_2\mathcal{F}(\mathbf{d}_1)^* \cdot \mathcal{F}(A)^* \cdot \mathcal{F}(A) \cdot \mathcal{F}(\mathbf{d}_1) \\ +\rho_2\mathcal{F}(\mathbf{d}_2)^* \cdot \mathcal{F}(A)^* \cdot \mathcal{F}(A) \cdot \mathcal{F}(\mathbf{d}_2) \\ +\rho_3\mathcal{F}(\mathbf{d}_3)^* \cdot \mathcal{F}(A)^* \cdot \mathcal{F}(A) \cdot \mathcal{F}(\mathbf{d}_3) + \|A\|^2]\},\end{aligned} \tag{16}$$

where $\mathcal{F}$ and $\mathcal{F}^{-1}$ compute the 3D fast Fourier transform and its inverse, respectively. The symbol ‘ $^*$’ denotes the complex conjugate, and ‘$/$’ and ‘$\cdot$’ indicate the element-wise division and multiplication, respectively.

### 2.3.2. MF-Decon with spatial RED and temporal TV

In the case of MF-Decon with spatial RED and temporal TV (MF-Decon+RED+TV), problem (6) is equivalent to:

$$\begin{aligned}\widehat{\mathbf{X}} = \underset{\mathbf{X}}{\operatorname{argmin}}\frac{1}{2}\|\mathbf{Y} - A * \mathbf{X}\|_F^2 + \mathcal{I}_+(\mathbf{Z}_1) + \lambda_1\|\mathbf{Z}_1\|_1 \\ +\lambda_2 RED(\mathbf{Z_2}) + \lambda_3\|\mathbf{Z}_4\|_1 \\ \text{s.t.}\quad \mathbf{Z}_1 = \mathbf{X}, \mathbf{Z_2} = A * \mathbf{X}, \mathbf{Z}_4 = \mathbf{d}_3 * (A * \mathbf{X}).\end{aligned} \tag{17}$$

Similarly, Algorithm 2 demonstrates the optimisation scheme for (17) with ILF-ADMM. According to (Romano et al., 2017), the proximal mapping associated with the RED regularisation can be computed iteratively, as follows:

$$\operatorname{prox}_{RED,\lambda_2,\rho_2}^{(m)}(x) = v^{(m)} = \frac{1}{\rho_2 + \lambda_2}\left(\rho_2 x + \lambda_2 f\left(v^{(m-1)}\right)\right), \tag{18}$$

where $f(\cdot)$ is an arbitrary image denoiser. The variable $v^{(m)}$ is introduced to simplify the notation, representing the value of the proximal operator at the $m$-th iteration. In this case, $\mathbf{X}^{(m)}$ can also be updated following:

$$(\rho_1\boldsymbol{\delta} + \rho_2 A' * A + \rho_3\mathbf{d}_3' * A' * A * \mathbf{d}_3 + \mathbf{h}) * \mathbf{X}^{(m)} = \mathbf{G}^{(m)}, \tag{19}$$

with

**Algorithm 2:** MF-Decon with spatial RED and temporal TV (MF-Decon+RED+TV) solved with ILF-ADMM

**Input:** Multi-frame CEUS images $\mathbf{Y}$ and estimated PSF $A$

**Output:** High-resolution deconvolved images $\widehat{\mathbf{X}}$

Initialise matrices $\mathbf{X}^{(0)} = \mathbf{0}$, ${\mathbf{Z}_1}^{(0)} = \mathbf{0}$, ${\mathbf{Z}_2}^{(0)} = \mathbf{0}$, ${\mathbf{Z}_4}^{(0)} = \mathbf{0}$, ${\tilde{\mathbf{Z}}_1}^{(0)} = \mathbf{0}$, ${\tilde{\mathbf{Z}}_2}^{(0)} = \mathbf{0}$, ${\tilde{\mathbf{Z}}_4}^{(0)} = \mathbf{0}$

$m = 1$

**while** *not converged* **do**

  ${\hat{\mathbf{Z}}_1}^{(m)} = \mathrm{prox}_{\mathcal{J}_+}\left(\mathbf{X}^{(m-1)} + {\tilde{\mathbf{Z}}_1}^{(m-1)}\right)$

  ${\mathbf{Z}_1}^{(m)} = \mathrm{prox}_{\|\cdot\|_{1,\lambda_1}}\left({\hat{\mathbf{Z}}_1}^{(m)}\right)$

  ${\mathbf{Z}_2}^{(m)} = \mathrm{prox}^{(m)}_{RED,\lambda_2,\rho_2}\left(A * \mathbf{X}^{(m-1)} + {\tilde{\mathbf{Z}}_2}^{(m-1)}\right)$

  ${\mathbf{Z}_4}^{(m)} = \mathrm{prox}_{\|\cdot\|_{1,\lambda_3}}\left(\mathbf{d}_3 * \left(A * \mathbf{X}^{(m-1)}\right) + {\tilde{\mathbf{Z}}_4}^{(m-1)}\right)$

  Update $\mathbf{X}^{(m)}$ according to (19)

  ${\tilde{\mathbf{Z}}_1}^{(m)} = {\tilde{\mathbf{Z}}_1}^{(m-1)} + \mathbf{X}^{(m)} - {\mathbf{Z}_1}^{(m)}$

  ${\tilde{\mathbf{Z}}_2}^{(m)} = {\tilde{\mathbf{Z}}_2}^{(m-1)} + A * \mathbf{X}^{(m)} - {\mathbf{Z}_2}^{(m)}$

  ${\tilde{\mathbf{Z}}_4}^{(m)} = {\tilde{\mathbf{Z}}_4}^{(m-1)} + \mathbf{d}_3 * \left(A * \mathbf{X}^{(m)}\right) - {\mathbf{Z}_4}^{(m)}$

  $m = m + 1$

$\widehat{\mathbf{X}} = \mathbf{X}^{(m)}$

$$\begin{aligned} \mathbf{G}^{(m)} = A' * \mathbf{Y} + \rho_1\left({\mathbf{Z}_1}^{(m)} - {\tilde{\mathbf{Z}}_1}^{(m-1)}\right) + \rho_2 A' * \left({\mathbf{Z}_2}^{(m)} - {\tilde{\mathbf{Z}}_2}^{(m-1)}\right) \\ +\rho_3 \mathbf{d}_3' * \left(A' * \left({\mathbf{Z}_4}^{(m)} - {\tilde{\mathbf{Z}}_4}^{(m-1)}\right)\right) + (\mathbf{h} - A' * A) * \mathbf{X}^{(m-1)}, \end{aligned} \quad (20)$$

which can also be solved in the Fourier domain as:

$$\begin{aligned} \mathbf{X}^{(m)} = \mathcal{F}^{-1}\{\mathcal{F}(\mathbf{G}^{(m)}) / [\,\rho_1 + \rho_2 \mathcal{F}(A)^* \cdot \mathcal{F}(A) \\ +\rho_3 \mathcal{F}(\mathbf{d}_3)^* \cdot \mathcal{F}(A)^* \cdot \mathcal{F}(A) \cdot \mathcal{F}(\mathbf{d}_3) + \|A\|^2]\}. \end{aligned} \quad (21)$$

## 2.4. Evaluations

### 2.4.1. In silico simulation

Our proposed MF-Decon methods, MF-Decon+3DTV and MF-Decon+RED-TV, are first evaluated with the simulated dataset generated with BUbble Flow Field (BUFF) (Lerendegui et al., 2022), where ground truth is available. This dataset was used in the Ultrasound Localisation and Tracking Algorithms for Super Resolution (ULTRA-SR) Challenge (Lerendegui et al., 2024) at the 2022 IEEE International Ultrasonics Symposium (IUS) conference.

In this dataset, a total of 500 frames were generated using the parameters of a high-frequency linear array L11-4 transducer with a centre frequency of 7.24 MHz, 3-angle compounding plane-wave imaging (angle step of 10˚), and a post compounding frame rate of 50 Hz. Noise is generated by filtering white Gaussian noise with the transducer bandwidth, after which it is added to the radio frequency (RF) channel signals. Full simulation parameters can be found in (Lerendegui et al., 2024).

Since the simulated data comes with the ground truth (GT), we can use the following three evaluation metrics as described in (Lerendegui et al., 2022, 2024):

$$\text{Recall} = \frac{\text{TP}}{\text{TP} + \text{FN}} \tag{22a}$$

$$\text{Precision} = \frac{\text{TP}}{\text{TP} + \text{FP}} \tag{22b}$$

$$F_1 = \frac{2\text{TP}}{2\text{TP} + \text{FN} + \text{FP}} \tag{22c}$$

where True Positive (TP) represents a case where a MB is localised within half a wavelength to a ground truth location; False Positive (FP) indicates a situation where a detected MB is not real; and the failure to detect a real MB is a False Negative (FN). Recall indicates the percentage of MBs that have been correctly localised out of all actual MBs, Precision tells the percentage of correctly localised MBs out of all the localised MBs, and $F_1$ score combines Recall and Precision into a single metric, balancing their trade-off. Moreover, the mean error ($\bar{E}$) and standard error ($E_\sigma$) between the detected positions ($P_{d,n}$) and the ground truth ones ($P_{gt,n}$) are also used to demonstrate the localisation accuracy, as defined in (Lerendegui et al., 2022):

$$\bar{E} = \frac{\sum_{n=1}^{N} \text{dist}\left(P_{d,n}, P_{gt,n}\right)}{N}, \tag{23a}$$

$$E_\sigma = \sqrt{\frac{\sum_{n=1}^{N}\left(\text{dist}\left(P_{d,n}, P_{gt,n}\right) - \bar{E}\right)^2}{N}}, \tag{23b}$$

where $\text{dist}\left(P_{d,n}, P_{gt,n}\right)$ calculates the distance between two positions.

#### 2.4.2. In silico simulation

Our proposed MF-Decon methods are also evaluated using the rat brain dataset from (Shin et al., 2024). A total of 20,000 frames from 80 data acquisitions were collected over 80 seconds with a high-frequency linear array transducer (L22-14vX, Verasonics Inc., Kirkland, WA) connected to a Vantage 256 system. A 5-angle compounding plane-wave imaging sequence (angle step of 1°) was used, with a centre frequency of 15.625 MHz, a pulse repetition frequency (PRF) of 28.57 kHz, and a post-compounding frame rate of 1000 Hz. To get the CEUS images from the data, singular value decomposition (SVD) was applied to remove the slow-moving tissue signals.

To make a fair comparison between our proposed MF-Decon and the state-of-the-art MB localisation methods, the same number of MBs were localised using different methods. They were then passed to the tracking algorithm in (Yan et al., 2022) with the same settings to generate SR microvasculature maps.

# 3. Results

## 3.1. Implementation details

All methods were implemented in MATLAB 2023b on a computer with an AMD Ryzen 9 5950X CPU and NVIDIA GeForce RTX 4090 GPU. Hyperparameters of the proposed methods were chosen empirically as follows: loss weights $\lambda_1 = 0.1$, $\lambda_3 = 2$ and $\lambda_2 = 0.1$ for MF-Decon+3DTV or $\lambda_2 = 2$ for MF-Decon+RED+TV; importance coefficients $\rho_1 = 10$, $\rho_3 = 0.1$ and $\rho_2 = 0.1$ for MF-Decon+3DTV or $\rho_2 = 1$ for MF-Decon+RED+TV. Furthermore, 500 iterations with a learning rate $\alpha = 20$ were found to be sufficient for all deconvolution-based methods to perform effective MB localisation. The denoiser used for RED was a median filter of size $5 \times 5$.

## 3.2. In silico simulation

We first evaluated the localisation performance of our proposed multi-frame methods (MF-Decon+3DTV and MF-Decon+RED+TV) on the in silico dataset, compared with the state-of-the-art methods: normalised cross-correlation (NCC) and traditional deconvolution (Decon).

Fig. 2 shows the precision-recall curves of different methods generated by changing the decision threshold after cross-correlation or deconvolution under three different SNR scenarios (SNR=5dB, 10dB and 15dB). It can be seen that the precision-recall curves of our proposed MF-Decon+3DTV and MF-Decon+RED+TV methods are consistently above those of NCC and Decon under a given SNR scenario, indicating more MBs can be accurately localised. More importantly, by changing the noise levels in the dataset from SNR=15dB to 5dB, we find that the noisier the data, the more the localisation results benefit from the proposed MF-Decon methods. Specifically, MF-Decon+3DTV can localise up to 12% more real MBs than NCC and 5% more than Decon, given the same total number of localised MBs; and it can localise MBs up to 39% more accurately than NCC and 15% more accurately than Decon, given the same number of real MBs being localised. Moreover, MF-Decon+RED+TV can localise up to 10% more real MBs than NCC and 3% more than Decon, given the same total number of localised MBs; while it can achieve up to 32% higher localisation accuracy than NCC and 8% higher accuracy than Decon when localising the same number of real MBs.

The best localisation results that each method can achieve, with respect to $F_1$ score, are summarised in Table 1. It can be observed that our proposed MF-Decon+3DTV and MF-Decon+RED+TV methods can always achieve higher $F_1$ scores than NCC and Decon methods, suggesting that our proposed methods can localise MBs more accurately and find more real MBs at the same time compared with the state-of-the-art methods. Furthermore, lower mean and standard location errors indicate that our proposed MF-Decon+3DTV and MF-Decon+RED+TV methods can also localise MBs closer to their ground-truth locations, compared with NCC and Decon methods. We also observe that the improvement in localisation performance by our proposed Decon+3DTV and MF-Decon+RED+TV methods increases as the noise level in the dataset decreases from SNR=15dB to 5dB.

To visually compare the localisation performance, we plot the super-localised MB maps generated by different methods with the same number of MBs detected under SNR=5dB against the ground truth (GT) map in Fig. 3, where MB intensities are normalised. In general, the maps generated with our MF-Decon+3DTV and MF-Decon+RED+TV methods have less background noise and are of higher contrast than those generated with NCC and Decon methods. Improvements by our Decon+3DTV and MF-Decon+RED+TV methods are more obvious when zooming in to the region highlighted in the green box. Our methods produce localisation maps that are more consistent with the GT map than other methods, as highlighted in the two green circles. Especially, an arc curve highlighted by a green arrow has been successfully reconstructed by our MF-Decon+3DTV and MF-Decon+RED+TV methods whereas it is less discernible in the maps reconstructed by NCC and Decon methods. Altogether the results on the in silico dataset demonstrate that our proposed MF-Decon+3DTV and MF-Decon+RED+TV methods can handle noise better and perform better MB localisation than the state-of-the-art methods.

## 3.3. In vivo rat brain data

Results tested on the rat brain data are shown in this section to demonstrate that our proposed MF-Decon+3DTV and MF-Decon+RED+TV methods not only work on simulations but also outperform the state-of-the-art methods on in vivo data.

Fig. 4 shows the maximum intensity projection (MIP) of a single short acquisition of the rat brain dataset, of 250 frames in total, and the super-localised MB maps generated by different methods, with the same number of MBs detected. The MB intensities are also normalised for a fair comparison. Although no GT is available, we can see that the maps generated with our MF-Decon+3DTV and MF-Decon+RED+TV methods exhibit less noise than those generated with NCC and Decon methods, as indicated by yellow circles and green arrows, respectively. Furthermore, zoom-in results in cyan and green boxes show that MBs localised by our MF-Decon+3DTV and MF-Decon+RED+TV methods form clearer and more distinct trajectories that are consistent with the structures in the MIP, in contrast to NCC and Decon methods, especially as highlighted by yellow arrows.

The SR maps of super-localised MBs generated with different methods on the whole in vivo rat brain dataset (of 20000 frames in total) are displayed in Fig. 5. Similarly, for a fair comparison, the maps are generated with same number of MBs localised and the bubble intensities are normalised. It is obvious that the SR maps produced with our MF-Decon+3DTV and MF-Decon+RED+TV methods contain more vessel structures than the one produced with NCC method. Compared to Decon method, the SR maps produced with our methods are less noisy and of higher contrast. For a better visual comparison, we focus on the results in the regions highlighted by cyan and green boxes, shown in Fig. 5(e) and Fig. 5(g), respectively. These figures illustrate that the microvasculature structures reconstructed by our MF-Decon+3DTV and MF-Decon+RED+TV methods are clearer and more visible, particularly when two vessels are very close to each other, as indicated by the yellow arrows. The MB intensity curves sampled along the cyan and green cross-section lines, plotted in Fig. 5(f) and Fig. 5(h), further demonstrate that our proposed methods can better separate MBs in adjacent vessels.

Moreover, the SR direction maps of super-localised MBs generated with different methods are also demonstrated in Fig. 6, where detected MBs moving upward are shown in orange and those moving downward are in cyan. In this way, vessels with blood flowing in different directions can be separated. From Fig. 6(a)-6(d), the same conclusion can be drawn that SR direction maps produced by our methods are less noisy, of higher contrast and with more vessel structures (as indicated by red arrows). We also focus on the zoom-in results in two regions highlighted by green and cyan boxes, shown in Fig. 6(e) and Fig. 6(f). From these figures, it is evident that the MB tracks generated by our MF-Decon+3DTV and MF-Decon+RED+TV methods are clearer and more distinct than those by NCC and Decon methods, leading to SR maps with more visible microvasculature structures and higher resolutions, as indicated by red arrows.

To further evaluate our proposed methods, box charts in Fig. 7 show the averaged length of the MB trajectories from each acquisition linked by the same tracking algorithm described in (Yan et al., 2022), based on same number of MBs localised by the different methods. Since the data in each group follows a normal distribution, a One-way ANOVA was performed on the four groups, giving $p$-value $= 4.87 \times 10^{-149}$. This was followed by Tukey's Honest Significant Difference (HSD) post hoc test, showing highly significant differences between all pairs of groups (all $p$-values $< 0.001$). Thus, the averaged lengths of the MB trajectories tracked based on our MF-Decon+3DTV and MF-Decon+RED+TV methods are significantly longer than those tracked based on NCC and Decon methods. In particular, our MF-Decon+RED+TV and MF-Decon+3DTV methods enhance the means of the averaged track lengths to 8.90 and 6.98 frames, representing increases of more than 102% and 59%, respectively, compared with NCC and Decon methods with the means of the averaged track lengths of 3.91 and 4.40 frames.

# 4. Discussion

SRUS imaging can be greatly affected by the noise in the acquired ultrasound data, creating extra challenges in localising the MBs from the CEUS images and thus in generating high-quality SR images. In this paper, we propose two MB localisation methods, MF-Decon+3DTV and MF-Decon+RED+TV, that can use the spatiotemporal coherence in the data to enhance the resolution and quality of the SR microvasculature maps. Our methods are based on a novel multi-frame deconvolution framework with spatial and temporal regularisers.

Numerical results from the in silico simulation show that our proposed MF-Decon+3DTV and MF-Decon+RED+TV methods enhance MB localisation performance across all evaluation matrices, including precision, recall, $F_1$ score, mean location error and standard error. More MBs are localised, and they are closer to the ground truth, compared with the state-of-the-art methods. With the in vivo rat brain dataset, we also demonstrate that the SR microvasculature maps generated with our methods exhibit less noise, better contrast, higher resolution and more vessel structures. In addition, we show that our methods enable longer MB trajectories to be tracked by the same tracking algorithm, compared with NCC and Decon methods.

The main feature of our methods, compared to the NCC and Decon ones, is that we leverage the spatiotemporal coherence of the data through spatial and temporal regularisers. The spatial regulariser based on 2D TV or RED suppresses noise in the spatial dimensions of the reconstructed images, thereby reducing the localisation errors caused by noise. More importantly, the temporal regulariser based on 1D TV imposes smoothness in the temporal dimension of the reconstructed data. The primary benefit of the temporal regulariser is its ability to further reduce noise, assuming that noise lacks temporal coherence and typically varies much faster than MB movements. With the temporal regulariser, we can eliminate fast-fluctuating noise but still keep the MB signals in the temporal dimension, which also prevents false localisation of MBs due to high-intensity noise. Secondly, it can reduce the localisation errors caused by noise and the shape variations of a MB in different frames. This allows for more accurate and consistent MB localisation across frames, which also helps the tracking algorithm to produce longer MB trajectories, resulting in more detailed vessel structures in the final SR maps.

A more important advantage of our proposed MF-Decon framework with RED (MF-Decon+RED+TV) is its ability to integrate various image denoising techniques into the deconvolution process. This offers substantial potential for enhancing the SR microvasculature maps by exploring different denoising approaches within the framework of RED. These include not only the traditional denoising filters that are commonly used in the post-processing pipelines to generate SR ultrasound images, for instance, medium filter, Wiener filter (Riemer et al., 2023), non-local mean filter (Song et al., 2018) and BM3D filter (Lei et al., 2022), but also the powerful denoising approaches that have been recently developed based on deep-learning, for example, TNRD (Chen and Pock, 2017), DnCNN (Zhang et al., 2017), DRUNet (Zhang et al., 2022) and so on. In fact, the denoising technique does not have to be limited to the two spatial dimensions. Three-dimensional denoisers can also be applied to further exploit the temporal coherence of the data. Moreover, potential improvement may also be achieved with TV of higher orders, due to their ability to enhance complex feature preservation, reduce artifacts, and improve noise reduction at the cost of increased computational complexity.

The proposed MF-Decon methods have the limitation of requiring more processing time compared to NCC and Decon methods, due to the addition of regularisers to the optimisation problem. To compensate for this, we applied a variant version of ADMM (ILF-ADMM) to solve the optimisations in this paper, which has demonstrated faster convergence. Moreover, since our proposed methods are mainly based on tensor convolutions and multiplications, they can be efficiently implemented on the GPU in MATLAB to achieve faster computational speeds. Under the configuration of an AMD Ryzen 9 5950X CPU and NVIDIA GeForce RTX 4090 GPU, our proposed MF-Decon+3DTV and MF-Decon+RED+TV methods need 9.35 minutes and 8.24 minutes, respectively, to deconvolve data consisting of 450 pixels by 650 pixels by 250 frames, compared to 2.29 minutes for Decon method and 4.82 minutes for NCC method (implemented on CPU). The proposed method also generated some artefacts at the boundaries of the imaged region. This issue arises because the shapes of MBs are cropped at the boundaries, leading to inaccurate localisation. However, since our regions of interest are normally in the middle of the ultrasound images, this problem can be easily solved by cropping the boundaries.

# 5. Conclusion

In this paper, we proposed two multi-frame deconvolution methods, MF-Decon+3DTV and MF-Decon+RED+TV, with spatial and temporal regularisers to exploit the spatiotemporal coherence of the acquired CEUS data, producing enhanced MB localisation performance for SRUS imaging. Our methods were evaluated on both in silico and in vivo datasets and the results verified that they outperform two widely used methods based on normalised cross-correlation and deconvolution. Consequently, the SR microvasculature maps generated with our methods exhibit less noise, better contrast, higher resolution and more vessel structures.

# Acknowledgements

This work was supported in part by the Biotechnology and Biological Sciences Research Council (BBSRC) under Grant BB/W001497/1, Grant BB/X017273/1, Grant BB/X011054/1, and Grant BB/N016947/1; in part by Diabetes UK (DUK) under Grant DUK 18/0005886 and Grant DUK 20/0006295; in part by Medical Research Council (MRC) under Grant MR/Y013980/1, in part by Engineering and Physical Sciences Research Council (EPSRC) Centre of Doctoral Training (CDT); in part by Chan Zuckerberg Foundation under Grant No. 2020-225443; and in part by Heptares.

Table 1: The best localisation results, in terms of $F_1$ scores, that each method performs on the in silico dataset when SNR=15dB, 10dB, and 5dB, where the best results are in bold and the second best underlined.

| SNR | Method | Precision | Recall | Mean error (μm) | Standard error (μm) | $F_1$ score |
|---|---|---|---|---|---|---|
| 15dB | NCC | 0.661 | 0.412 | 45.278 | 25.669 | 0.508 |
| | Decon | 0.859 | 0.557 | 37.472 | 23.417 | 0.676 |
| | MF-Decon+3DTV | **0.890** | **0.566** | <u>37.274</u> | <u>23.368</u> | **0.692** |
| | MF-Decon+RED+TV | <u>0.869</u> | **0.566** | **36.687** | **23.247** | <u>0.686</u> |
| 10dB | NCC | 0.577 | 0.325 | 49.049 | 26.051 | 0.416 |
| | Decon | 0.718 | 0.427 | 42.896 | 24.646 | 0.536 |
| | MF-Decon+3DTV | **0.777** | **0.442** | <u>42.337</u> | <u>24.622</u> | **0.563** |
| | MF-Decon+RED+TV | <u>0.730</u> | <u>0.439</u> | **42.171** | **24.565** | <u>0.549</u> |
| 5dB | NCC | 0.379 | 0.219 | 54.240 | 26.433 | 0.278 |
| | Decon | 0.451 | 0.282 | 49.666 | 25.869 | 0.347 |
| | MF-Decon+3DTV | **0.551** | <u>0.285</u> | **48.183** | **25.664** | **0.375** |
| | MF-Decon+RED+TV | <u>0.495</u> | **0.287** | <u>48.474</u> | <u>25.762</u> | <u>0.363</u> |

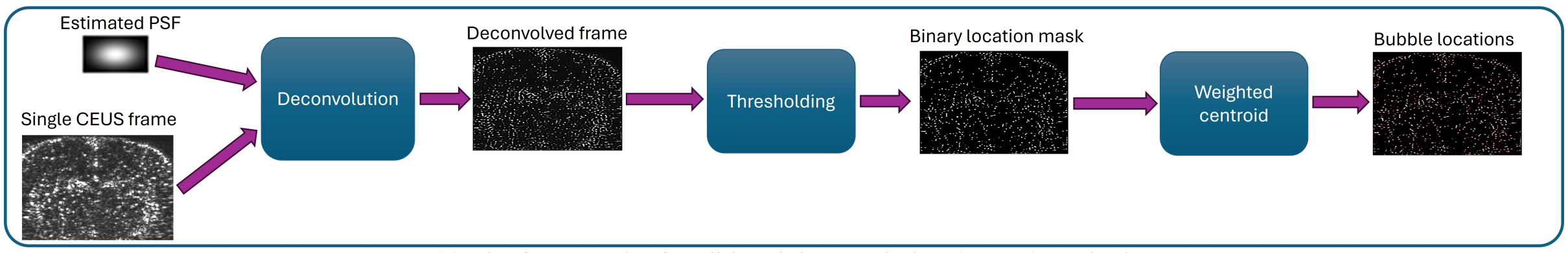

(a) The framework of traditional deconvolution (Decon) method

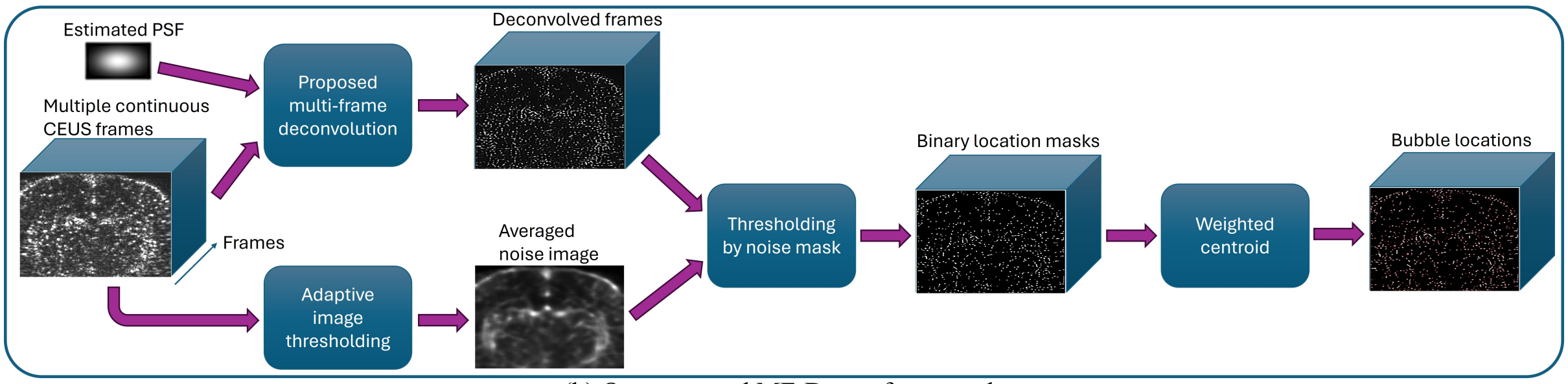

(b) Our proposed MF-Decon framework

Figure 1: The frameworks of (a) traditional deconvolution (Decon) and (b) our proposed Multi-Frame Deconvolution (MF-Decon). The Decon method processes the CEUS images frame by frame independently. However, the proposed MF-Decon framework takes multiple continuous CEUS frames as input, so that spatiotemporal coherence can be exploited during deconvolution.

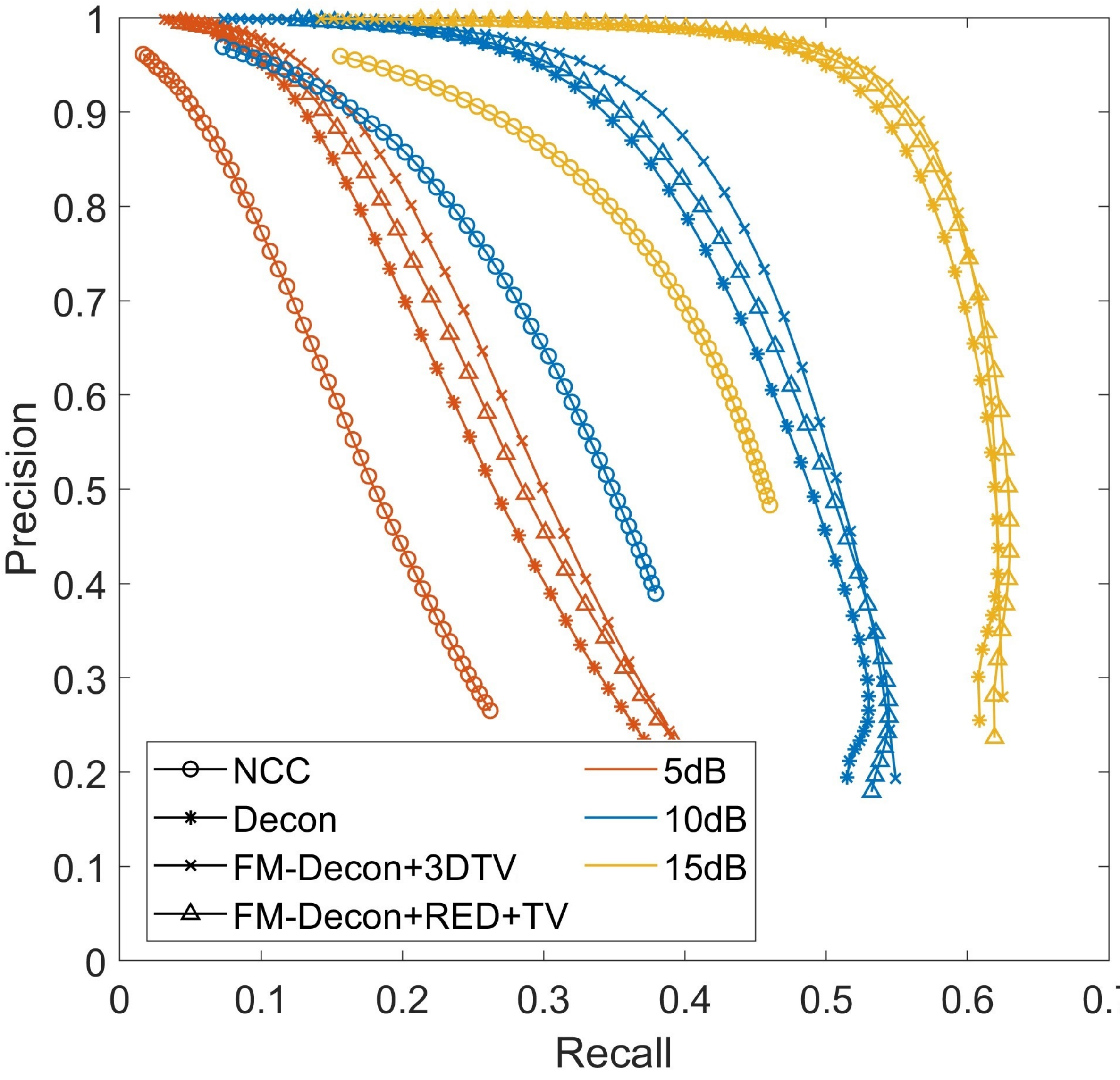

Figure 2: Precision and recall curves of different methods under three different SNR scenarios (SNR=5dB, 10dB and 15dB).

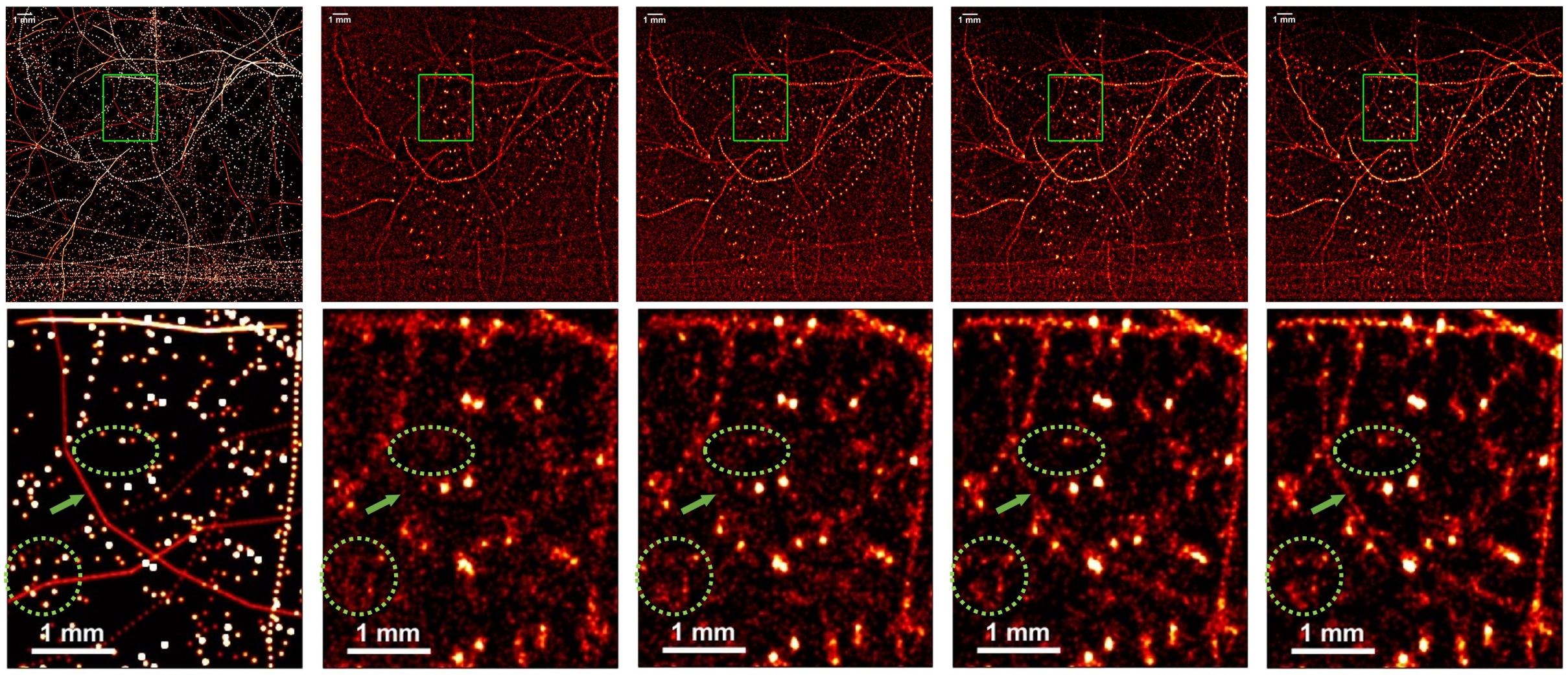

(a) GT (b) NCC (c) Decon (d) MF-Decon+3DTV (e) MF-Decon+RED+TV

Figure 3: Super-localised MB maps generated with different methods on the in silico dataset under SNR=5dB: (a) ground truth (GT), (b) NCC, (c) Decon, (d) MF-Decon+3DTV, (e) MF-Decon+RED+TV. The second row shows the zoomed-in results in the green-boxed region, where the green ovals and arrows highlight some improvements by our methods.

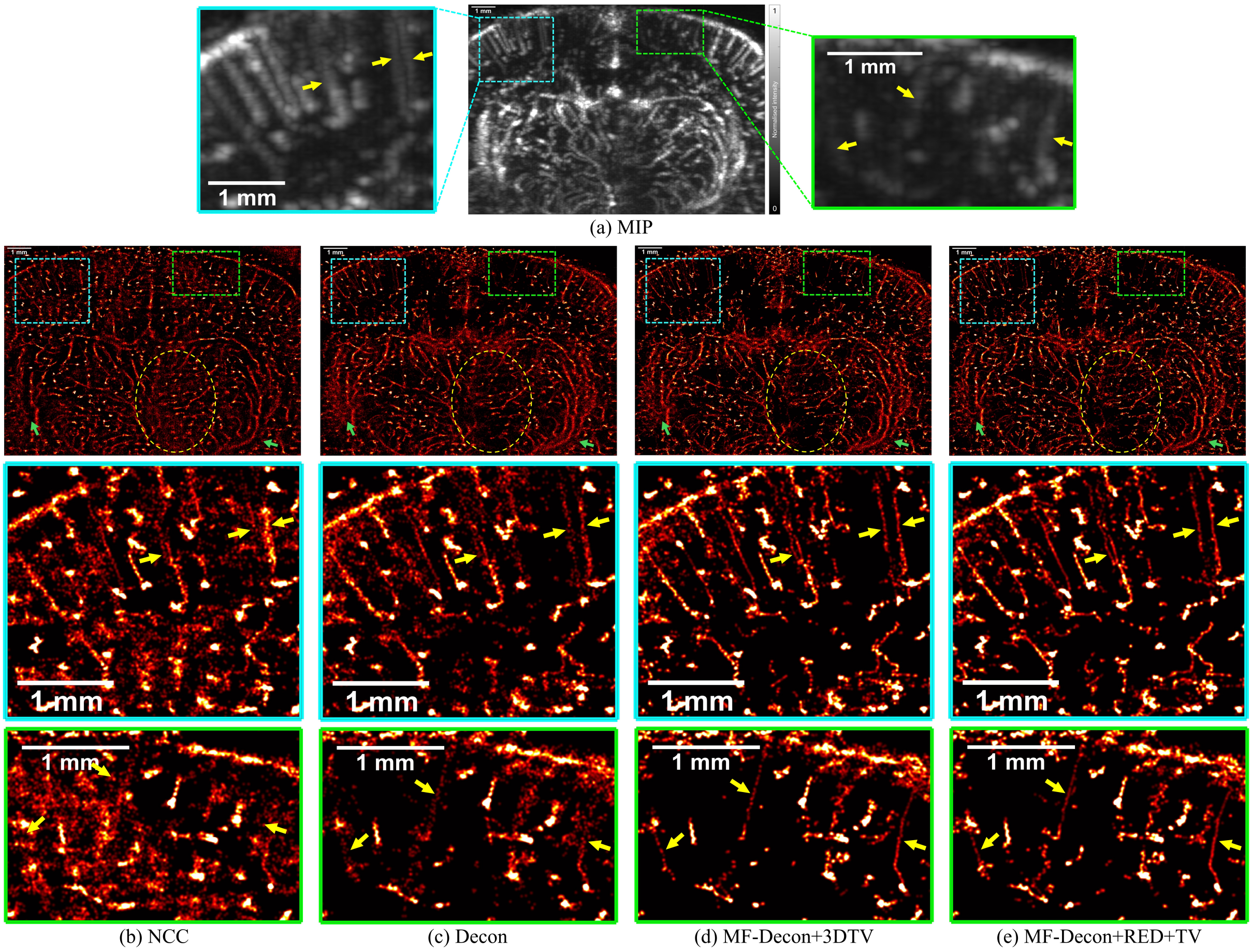

Figure 4: (a) shows maximum intensity projection (MIP) of a single acquisition (250 frames in total). (b)-(e) demonstrate super-resolved location maps generated on this acquisition with NCC, Decon, MF-Decon+3DTV and MF-Decon+RED+TV methods, respectively, where the zoomed-in results in the cyan-boxed and green-boxed regions are shown in the second and third rows, separately. Yellow ovals and green and yellow arrows highlight some improvements by our methods.

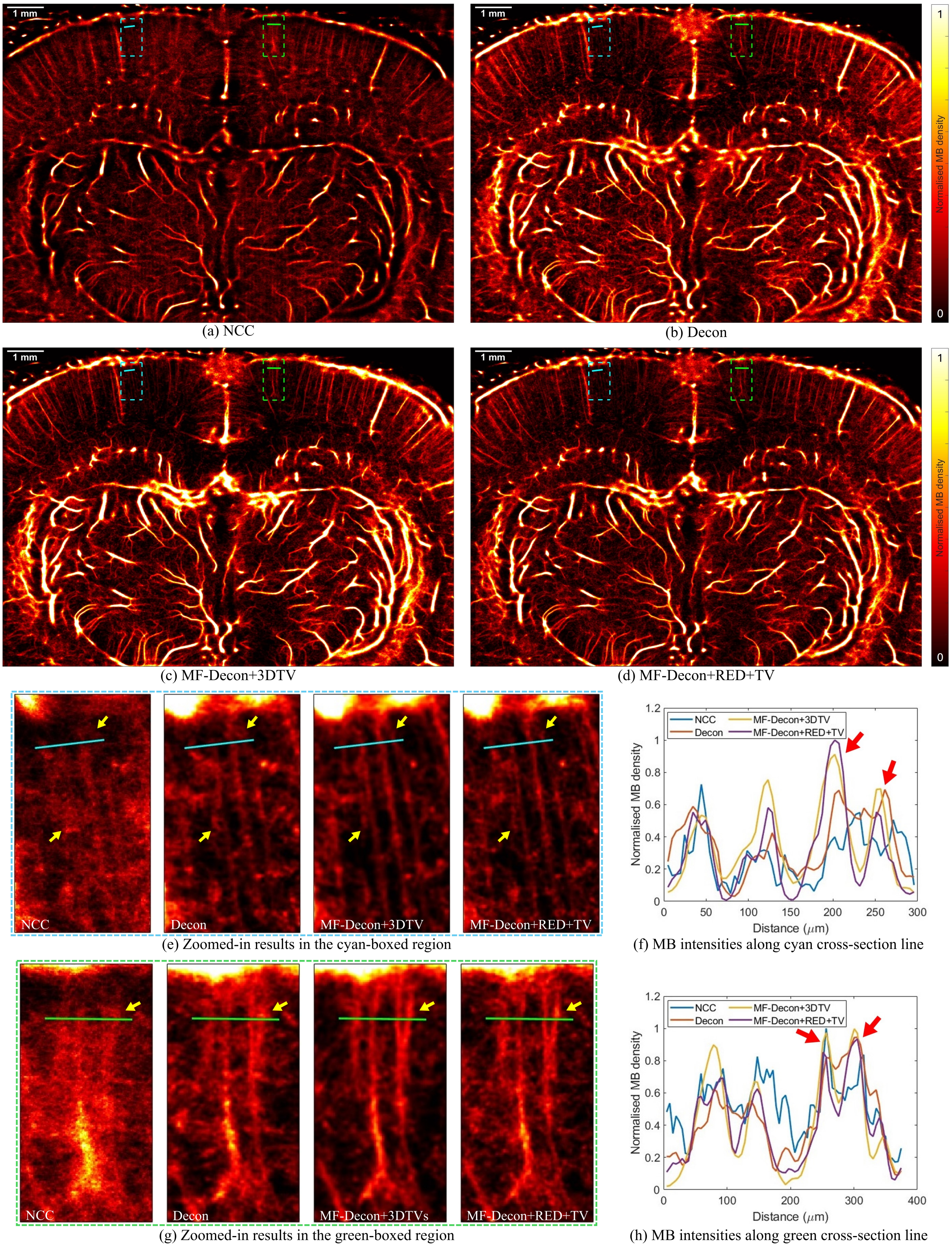

Figure 5: SR maps of super-localised MBs generated with different methods on the in vivo rat brain dataset. (a)-(d) generated with NCC, Decon, MF-Decon+3DTV and MF-Decon+RED+TV methods, respectively. (e) and (g) show zoomed-in results in the cyan-boxed and green-boxed regions. (f) and (h) show normalised intensities of super-localised MBs sampled along cyan and green cross-section lines. Yellow and red arrows highlight some improvements by our methods.

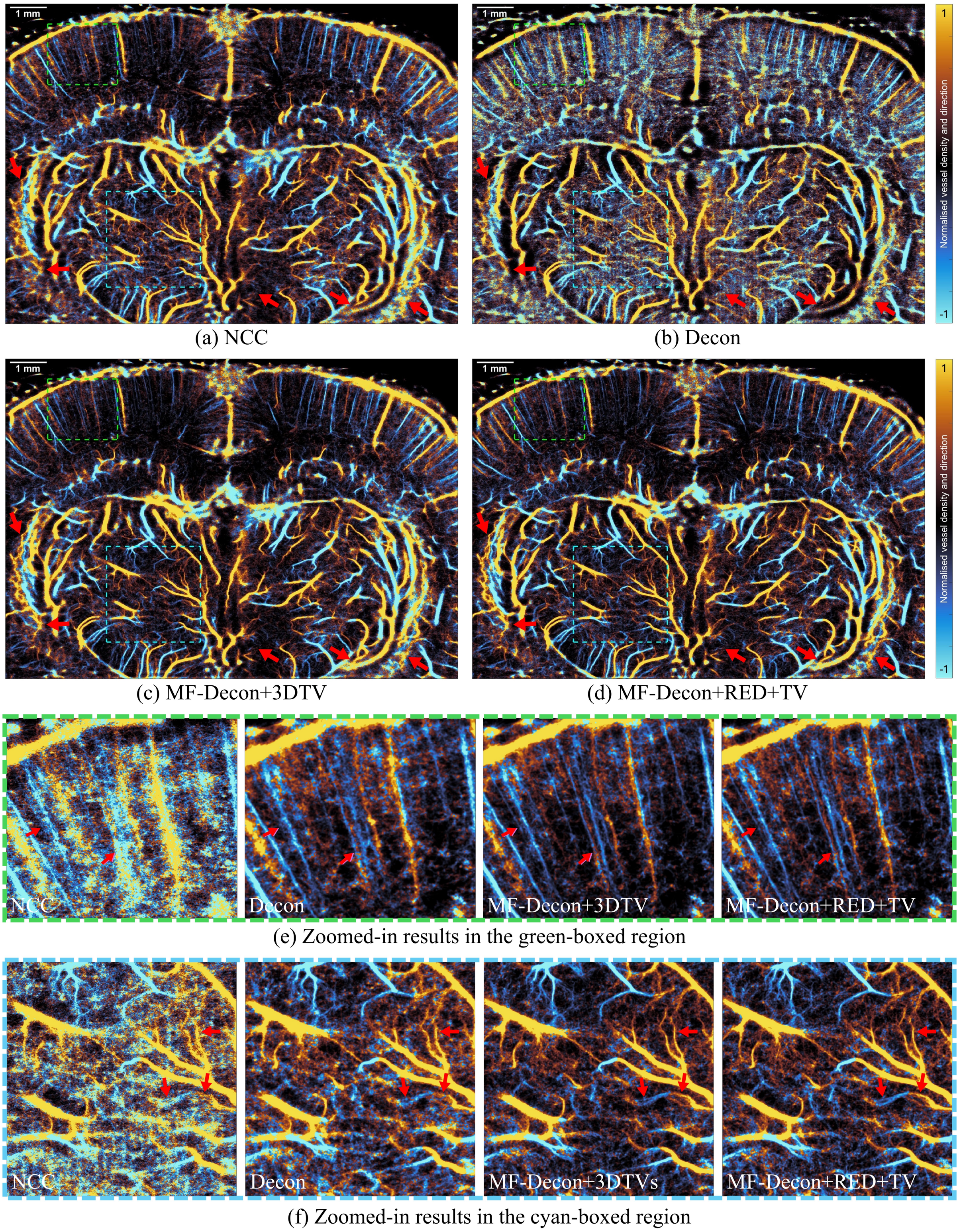

Figure 6: SR direction maps of super-localised MBs generated with different methods on the in vivo rat brain dataset. (a)-(d) generated with NCC, Decon, MF-Decon+3DTV and MF-Decon+RED+TV methods, respectively. (e) and (f) show zoomed-in results in the cyan-boxed and green-boxed regions. Red arrows highlight some improvements by our methods.

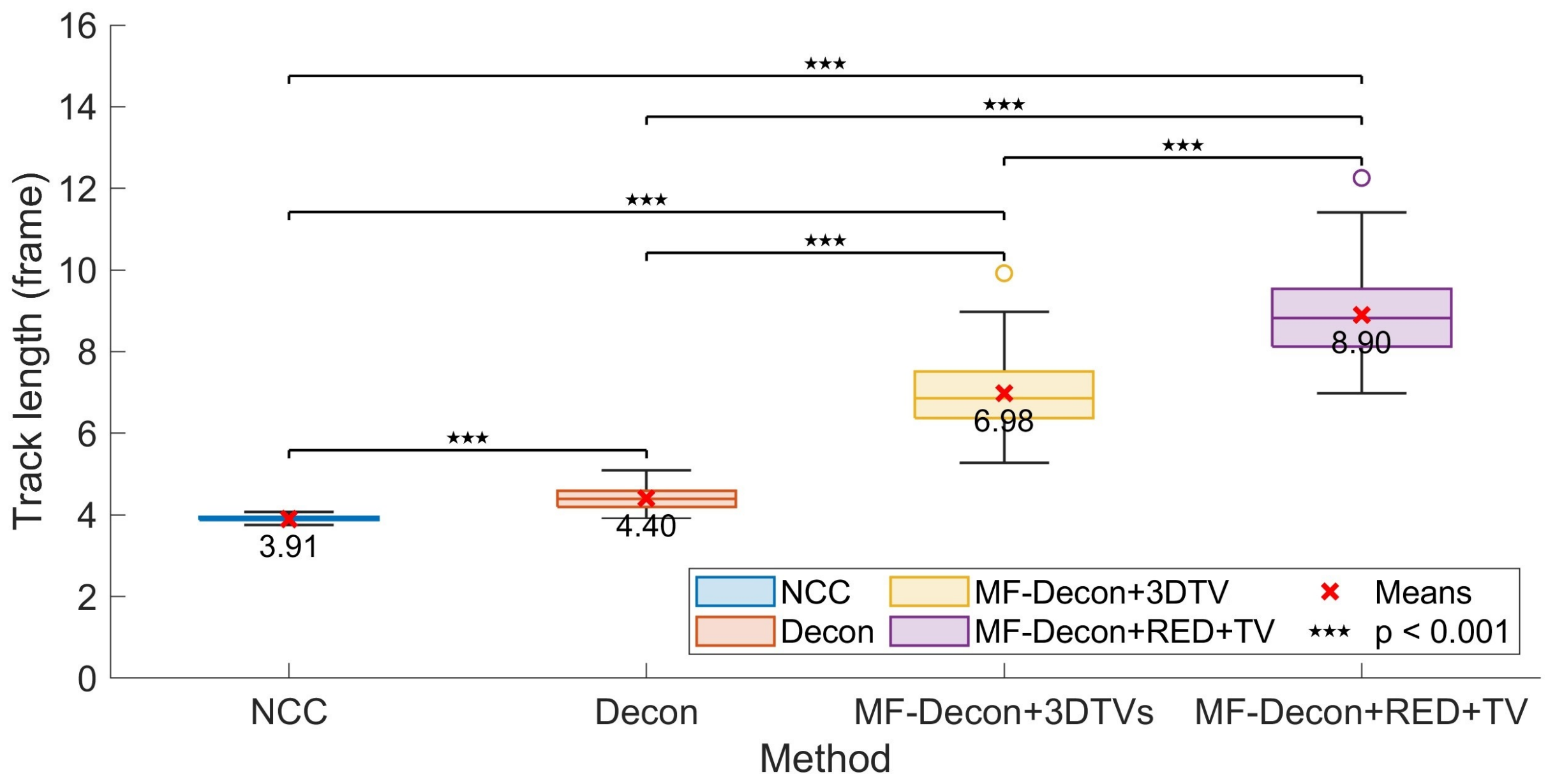

Figure 7: Box charts show the averaged track length of the MB trajectories from each acquisition linked by the same tracking algorithm, based on same number of MBs localised by different methods. Each box shows the median as the central line, the upper and lower quartiles as the top and bottom edges, outliers as circles, and the nonoutlier maximum and minimum as whiskers.